\documentstyle[12pt]{article}

\large
\newcommand{\be}{\begin{eqnarray}}
\newcommand{\ee}{\end{eqnarray}}

\newcommand\del{\partial}

\begin{document}
\setlength{\baselineskip}{21pt}
\pagestyle{empty}
\vfill
\eject
\begin{flushright}
SUNY-NTG-92/39
\end{flushright}

\vskip 2.0cm
\centerline{\bf Quark Propagation in the Random Instanton Vacuum}
\vskip 2.0 cm
\centerline{E.V. Shuryak and J.J.M.
Verbaarschot}
\vskip .2cm
\centerline{Department of Physics}
\centerline{SUNY, Stony Brook, New York 11794}
\vskip 2cm

\centerline{\bf Abstract}
  This is the first of a series of papers devoted to a systematic study
of QCD correlation functions in a framework of
'instanton vacuum' models.
The topic of this paper is to work out approximate formulae for quark
propagators in a multi-instanton environment.
As an application, and also as a necessary step toward
understanding the correlation functions, we
study  the propagators of scalar and spinor quarks, using
the simplest possible model, the so
called 'random instanton vacuum' (RIV). Results
related to heavy-light mesons, are found to be very consistent with
phenomenology.
\vfill
\noindent
\begin{flushleft}
SUNY-NTG-92/39\\
December 1992
\end{flushleft}
\eject
\pagestyle{plain}
\vskip 1cm
\centerline{\bf 1. Introduction}
\vskip .5cm

The understanding of non-perturbative phenomena in the QCD vacuum
remains to be a challenging problem of contemporary physics.
One particular way to obtain insights in the complicated
quark-quark and quark-anti-quark interactions is provided by
{\it  point-to-point correlation functions}, which,
at medium distances of 1/3-1 $fm$,  display strong
deviations from free quark propagation. These deviations are
strongly channel dependent and change widely in both sign and magnitude
(see \cite{SHURYAK-1992} for a recent review).

   Although the hadronic spectrum can be described  more or less
in terms of a universal confining
force and perturbative spin forces, such a simplified
picture definitely fails to describe the point-to-point
correlation functions. The actual situation is much more complex,
 and additional theoretical inputs are definitely needed.

Tunnelling phenomena in the QCD vacuum, described semiclassically by
the so called instanton solutions were suggested
as  explanations of at least some of these
effects (or even most of them). Classical papers of the
{\it one-instanton era} describe the
discovery of the instanton solution \cite{BELAVIN-ETAL-1975},
its physical interpretation as tunneling and the relation to chiral anomalies
\cite{THOOFT-1976A}, and the semiclassical integration of the quantum
fluctuations \cite{THOOFT-1976B}.
The first applications of instantons to QCD problems
\cite{CALLAN-DASHEN-GROSS-1978},
based on the so called {\it dilute gas approximation},
attracted a lot of attention  in the late seventies. However, the
absence of an explanation
for the diluteness and the semiclassical nature of the instanton ensemble
led to such a pessimism that most people left the field around 1980.

However, at the same time,
phenomenological studies of  instanton-related effects
\cite{GESHKENBEIN-IOFFE-1980,NOVIKOV-ETAL-1981}
have shown
that the induced effective  interaction of light quarks \cite{THOOFT-1976A}
has exactly the properties needed to
explain several puzzles of the hadronic world.   Eventually, this development
has resulted in the so called  'instanton   liquid' model
\cite{SHURYAK-1982,SHURYAK-1983D}, and its  phenomenological success
revived the hopes of describing the QCD vacuum by
a dilute system of semiclassical
fluctuations.

Multiple attempts \cite{CALLAN-DASHEN-GROSS-1979A,CARLITZ-CREAMER-1979}
\cite{ILGENFRITZ-MUELLER-PREUSSKER-1981,CARNEIRO-MCDOUGALL-1984}
\cite{DIAKONOV-PETROV-1984,DIAKONOV-PETROV-1986}
were made to create a quantitative theory
of the 'instanton component' of the QCD vacuum. The problem
appears to be very difficult: even if one disregards
all other non-perturbative phenomena and considers only instantons,
it is  complicated both by
the gauge-induced and, especially,
by the fermion-induced interaction between them.
The  theory of these phenomena,  which we
refer to as the {\it interacting instanton approximation}  (IIA)\footnote{For
a recent review and references on this subject we refer to
\cite{SHURYAK-1992}.},
has been developed during the past decade, and the present series paper
can be considered as an attempt to clarify its relation to experimental
information on correlators in a quantitative way.

   So far, the only numerical evaluation of correlators in a multi-instanton
environment was made in a series of works by
  one of us \cite{SHURYAK-1989I,SHURYAK-1989III}. The first results
on a set of correlation functions
 have well reproduced the data, sometimes in surprising detail. However,
the accuracy of the calculations was not high, and both the statistical  and
the systematic errors should be significantly reduced. This is one of the tasks
addressed by the present series of papers.

   Recently, the first results for point-to-point correlators
from lattice simulations \cite{CHU-ETAL-1992}
were reported. Remarkably, inside uncertainties
 they generally agree both with experiment and IIA predictions.
Whether these results are or are not related to instantons can be clarified by
further lattice studies. In the long run much better lattice data
will be obtained, so that a more detailed comparison with
experiment becomes possible. In any case,
this recent development makes the situation in the field
very exciting.

The present paper is the first in a series of three works. The
objective is to perform a {\it systematic} calculation of
 {\it all major} correlation functions
at distances of the order of $1\, fm$.
A systematic comparison of the results
with experimental  data is made whenever
possible. We also compare our results with predictions of the
operator product expansion (OPE), on which the so called
QCD sum rules are based.
As  this task happens to be quite formidable by itself, merely because
of the literally dozens of different correlation functions $etc.$,
we feel that pushing forward  the theory of interacting instantons
at the same time would be very confusing.
Instead, we have adopted the  idea to establish a kind of 'benchmark
calculation' for all correlators, using an as simple input model as possible.
The particular
framework used below will be called the 'random instanton vacuum' (RIV).
It is nothing else but the 'instanton liquid model', originally suggested
by one of us \cite{SHURYAK-1982},
supplemented by the simplest assumption of a random distribution of the
collective coordinates. We do not fit any parameters,
leaving them as they were proposed a decade ago at the one-instanton level.
It is relatively easy to
perform high-statistics numerical studies of this simple model, and, as we
demonstrate below and in the next papers of this series
\cite{SHURYAK-VERBAARSCHOT-1992D,SHURYAK-VERBAARSCHOT-1992E},
it  reproduces most of the experimental
observations amazingly well.

The present paper is organized as follows. In section 2
we discuss the main ingredients of our model, and we
pay particular attention to the zero mode sector of the Dirac operator.
Numerical results for its spectrum will be  presented. The general structure
of the quark propagator and its numerical evaluation will be discussed in
section 3. In section 4 we discuss the propagation of a scalar quark in the
RIV. The more complicated spinor propagator is discussed in section 5, and
the results are applied to heavy-light systems in section
6. Concluding remarks
are made in section 7.

The mesonic \cite{SHURYAK-VERBAARSCHOT-1992D}
and baryonic \cite{SHURYAK-VERBAARSCHOT-1992E} correlation functions are
discussed in two subsequent papers of this series.
After such benchmark for all correlators is established,
we are planning to return to their discussion in the presence
of {\it  instanton interactions}
with correlations among the positions and orientations.
Those  studyies are subject of future investigations.

\renewcommand{\theequation}{2.\arabic{equation}}
\setcounter{equation}{0}
\vskip 1.5cm
\centerline{\bf 2. The model and the zero mode zone}
\vskip 0.5cm

   The general reason why instantons are so important for the physics
of {\it light quarks} is related to the so called {\it
fermionic zero mode}, which exists for any topologically
non-trivial gauge field. In order to explain this let us consider
the Euclidean quark propagator defined by
\be S= - <{1 \over iD_\mu\gamma_\mu +im}>, \ee
where $iD_\mu= i\del_\mu +A_\mu$ is the  covariant derivative
containing the external gauge field,
which should be averaged over with the proper weight following from the QCD
partition function.
The operator may be inverted by first diagonalizing it,
\be
iD_\mu\gamma_\mu \phi_\lambda(x)=\lambda \phi_\lambda(x),
\ee
which leads to the general expression
\be S(x,y)=-\sum_{\lambda} {\phi_\lambda(x) \phi^\dagger_\lambda(y)
\over \lambda +im} \ee
for the propagator in the gauge field $A_\mu$.
It is clear that  in the chiral limit ($m \rightarrow 0$)
the {\it eigenvalue spectrum} at small $\lambda$
has important effects on the propagation of hadrons.
In particular, one can easily derive the
following general formula \cite{BANKS-CASHER-1980} for  the Euclidean
quark condensate
\be
<\bar\psi\psi>_E=\frac{\pi i}{V} \rho(\lambda=0),
\ee
where $\rho(\lambda)$ is the spectral density\footnote{We do not
repeat here the standard arguments on the order of the limits
$m\rightarrow 0$ and $V\rightarrow \infty$.
For a very detailed recent discussion of this topic
see \cite{LEUTWYLER-SMILGA-1992A}.}.

 'Zero modes',  discovered in \cite{THOOFT-1976A}, are solutions of
the Dirac equation in the field of {\it
one} instanton  with eigenvalue $\lambda=0$.
Their existence explains the so called $U(1)$
chiral anomaly: while tunneling,  quarks with one chirality
'dive into the Dirac sea', and those with the opposite chirality
'emerge' from it. In
\cite{SHURYAK-VERBAARSCHOT-1992D,SHURYAK-VERBAARSCHOT-1992E}
we will show that these phenomena are responsible for a
much wider range of physical effects than was
believed previously. In particularly, not only the
$\eta'$ channel is affected, but, say correlators in
the $\pi$ meson or the nucleon channels can be reproduced as well.

The  evaluation of a quark propagator in the multi-instanton
background field is a complicated problem.
In particular, even a
more or less {\it dilute} set
  of  instantons (and
anti-instantons) is described by  12 collective
variables per instanton (position, size and orientation),
and the spectrum of the Dirac operator depends on all of them. Roughly
speaking, the problem resembles the propagation of a sound wave
through
a turbulent atmosphere, containing many vortices and anti-vortices.

  To explain our approximations, one
can use an analogy from solid state physics in which the role of atoms
is played by instantons, and the electron bound state is represented by
the quark 'zero mode'.
The general  lesson from condensed matter physics is that
at finite density of the atoms, atomic  bound states may
become collective and form  'zones' of delocalized states\footnote{It
is important that such zones exist as well for disordered systems
(liquids), although in this case they do not have sharp boundaries.}.
  Analogously,  fermionic states in the 'instanton vacuum'
form the so called 'zero mode zone' (ZMZ)\footnote{What is important here, is
 that the mode for an individual 'well' is {\it
exactly at
zero}, due to some topological theorem: therefore  all small distortions are
irrelevant. We thus deal with a degenerate situation, in which
the spectrum is governed by the {\it off-diagonal} matrix elements.}.

 Furthermore, if the ensemble of instantons is dilute, the off-diagonal matrix
elements are small and the zone should
be narrow. If so, this set of states may dominate the spectrum
near zero  virtuality, and therefore
be mainly responsible  for $\rho(\lambda=0)$ and the quark condensate.
Direct  evidence that this indeed happens in the QCD vacuum has been
obtained in  lattice studies \cite{HANDS-TEPER-1990,CHU-HUANG-1991}. Much
more work in this direction can and
should be done in  order to clarify the relation between
lattice results and the theory of instantons.

  Returning to the propagator, one arrives at
a comprehensive physical picture of how light quarks can  easily
propagate over large distances
in the QCD vacuum: they can do it
by simply jumping from one instanton to the  next.
It is analogous to what conduction electrons do, in
  solid or liquid metals. Due to the degeneracy of atomic levels, quite small
perturbations can crucially enhance or reduce probability of this to happen.

  Continuing this analogy further,
 we have learned from condensed matter physics
that going beyond the one-electron approximation one comes across
an important interaction
between conduction electrons due to
 the possibility, that
two of them may happen to be {\it at the same atom}.
The same is true for quarks in
the instanton vacuum: the fact that they may meet each other at
{\it the same instanton} leads to their
strong interaction. Technically this interaction is revealed, when one compares
the average propagator {\it squared} (for mesons) or {\it cubed} (for
baryons) to the same power of the {\it average propagator}.
The latter is studied in detail below, while the former quantities will
 discussed in two subsequent
papers.

  Having outlined the main physics involved, let us now specify the ZMZ in more
detail. In a basis of zero modes \cite{DIAKONOV-PETROV-1986},
the Dirac operator
reduces to the following matrix
\be i\hat D=\left (
\begin{array}{cc}  0 & -T_{I \bar I}\\
                -T_{I \bar I}^{\dagger} & 0
\end{array} \right ),
\ee
where the overlap matrix elements are defined by
\be
T_{I\bar I} =\int d^4x \phi_I^*(x)i\gamma\cdot\del \phi_{\bar I}(x).
\ee
The ordinary derivative appears because the gauge field in the
covariant derivative has been eliminated by using the equations of
motion for the zero modes.
Although it is not very important for what follows,
in this paper we will use
the overlap matrix elements corresponding to 'streamline'
gauge field configurations.
They are discussed in
\cite{SHURYAK-VERBAARSCHOT-1992A} and are equal to
\be
T_{I \bar I } =
\frac 12 {\rm Tr} \left ( \frac{\sigma_\mu^+ R_\mu}{R} U_I^{-1}
U_{\bar I} \right )
\frac {F(\lambda)}{(\rho_I \rho_{\bar I})^{1/2}},
\ee
where $R = R_I - R_{\bar I}$ is four-distance between the centers,
$U_I^{-1} U_{\bar I} $ is the relative orientation of the instantons
and $\lambda$ is a conformally invariant parameter defined by
\be
\lambda  = \frac 12 \frac{R^2+ \rho_I^2 + \rho_{\bar I}^2}{\rho_I\rho_{\bar I}}
+ \frac 12\left  (\frac{(R^2 + \rho_I^2 + \rho_{\bar I}^2)^2}
{\rho_I^2 \rho_{\bar I}^2} -4\right)^{\frac 12},
\ee
Here and below we use the standard color matrices
$\sigma_\mu ^{\pm}=(\vec\sigma,\mp i)_\mu$. These matrices project the
relative orientation matrix onto its
upper $2\times 2$ block.
The scalar function $F(\lambda)$ can be reduced  to a simple one dimensional
integral
\be
 F(\lambda) =
6\int_0^\infty {dr r^{3/2} \over (r+1/\lambda)^{3/2} (r+\lambda)^{5/2}}.
\ee
Asymptotically, for large $R$ the overlap matrix elements behave as
$T \sim \rho^2/R^3$, and thus correspond
to the exchange of a massless quark.

   In this set of papers we are going to use
 a simplified instanton ensemble, the so called {\it random instanton
vacuum} (RIV). We assume the following: (i) all instantons
have the same size $\rho_0$; (ii) they have {\it random}
positions and orientations;
(iii) the instanton and anti-instanton densities are both equal to $N/2V$,
where $V$ is the volume of the Euclidean space time and $N$ is the
total number of pseudoparticles.

Thus, there is essentially only one dimensionless parameter in the model
$f=\pi^2 V/N \rho^4_0$ which describes 'diluteness'
of the vacuum. All propagators and correlators to be considered, normalized
with respect to their values for free massless quarks, are also given by
dimensionless ratios. Those can only
 depend  on $f$, while the overall {\it distance scale}
 is given by a second parameter, $e.g.$ by
 the 'average separation' $R$ defined by $(V/N)^{1/4}$.

As suggested a decade ago in \cite{SHURYAK-1982}, the parameters are chosen
such that several bulk properties of the QCD vacuum are reproduced.
In order to obtain the traditional value for gluonic condensate one has to take
$R = 1.0\, fm$ (inside uncertainties),
whereas the quark condensate (and several other phenomenological parameters)
are reproduced for $\rho_0=1/3 \,fm$.

  Somewhat surprisingly, already this simple model leads to a very reasonable
description of many correlation functions. This does not mean that we have
a full description of the QCD vacuum, and a number of disclaimers have
to be made:

(i) First of all,
the model does not include  neither the (so far
mysterious) field fluctuations leading to {\it confinement}, nor even
perturbative fields, leading to phenomena like the {\it Coulomb interaction}
or {\it radiative corrections} to 'asymptotically free' propagation.
The confining fields are supposed to modify our results substantially
for quarks travelling
far apart, while perturbative corrections are of
$O(\alpha_s(x)/\pi) \sim 10-20 \%$ at relevant distances.

(ii) The overall good performance of the model does not mean that {\it all}
results are good. For example, the model
definitely 'overshoots' the repulsion in the isovector scalar mesonic channel
\cite{SHURYAK-VERBAARSCHOT-1992D}, predicting
a negative correlator in some window of distances. It violates
positivity and is obviously wrong.

(iii) Another sector, in which
RIV does not perform correctly,
is related to large-scale fluctuations of the topological charge.
Obviously, a random distribution of instantons
implies that in any volume $V$ one expects to have excess of charge $\delta Q
\sim V^{1/2}$. In other words, the
 topological susceptibility $\chi = <Q^2>/V$ is {\it non-zero}.
 This feature is known to be wrong in the case of massless fermions,
where $\chi$ should vanish for large volumes. Thus, one expects that
the $\eta'$ correlator evaluated in the RIV 'overshoots'
the right behaviour, which is indeed what was found
\cite{SHURYAK-VERBAARSCHOT-1992D}.

  The strong fluctuations
of the topological charge can also be seen directly in the
spectrum of the Dirac operator and in the quark condensate.
The generic argument for the eigenvalue distribution goes as follows.
 If the limit
$N_c \rightarrow \infty$ is taken before the thermodynamic limit, the matrix
elements behave as independent random variables resulting in a semi-circular
eigenvalue distribution \cite{DIAKONOV-PETROV-1986,BRODY-ETAL-1981}.
However, if  the thermodynamic limit is taken at finite $N_c$, as it should,
the number of matrix elements is much larger than
the number of independent collective variables, leading to strong correlations
between them. In such case one expects the
resulting eigenvalue distribution to be
gaussian \cite{DIAKONOV-PETROV-1984,BRODY-ETAL-1981}.
On the other hand,  an excess $\delta Q$ of topological
charge  may create extra quasizero modes, but the naive argument\footnote{If
one ignores the interaction with the zero modes of the '$Q = 0$ background',
one finds for a topological charge excess of
$\delta Q \sim \sqrt V$ and corresponding overlap overlap matrix elements
of $T \sim \rho^2/V^{3/4}$ a level spacing of $\sim 1/V^{5/4}$ leading
to a divergent spectral density.} giving rise to a divergent spectral
level density $\rho(\lambda=0)$ disagrees with our numerical simulations
of the RIV. This is hardly surprising since the eigenvalues depend in a
very nonlinear way on the matrix elements.

Numerical results for an ensemble of 2560 configurations of 128
instantons (dotted line) and 160 configurations  of 512
instantons (full line)
are shown in Fig. 1. In both cases the density of instantons $N/V = 1$.
The two spectra nearly coincide and are indistinguishable
for small $\lambda$, even if the bin size is decreased by an order of
magnitude.
This suggests that the thermodynamic limit will not alter our results.
The dashed line is the corresponding gaussian distribution with
variance given by $\sigma^2 = 2 <{\rm Tr} (T T^\dagger)/N>$, where
$<\cdots>$ denotes ensemble averaging.
At small $\lambda$ our results
suggest a strong departure from the gaussian distribution, which is related
to the strong fluctuations of the topological charge. Although
the spectrum itself does not become infinite,
it seems to show an infinite slope $\rho'(\lambda)$ at $\lambda = 0$.
In terms of the normalized cumulants $\kappa_4/\kappa_2^2$ and
$\kappa_6/\kappa_2^3$ with values of 0.180 and $-0.910$, respectively,
the deviations from the gaussian distribution are much less visible.
We also have checked that a semi-circular distribution is obtained when
the overlap matrix elements are distributed as independent gaussian
random variables with zero mean.

The quark condensate, evaluated from the spectral density at $\lambda=0$,
is enhanced by a significant factor, and fluctuations
of the quark condensate, invoking the fluctuation-dissipation theorem,
 diverge in the chiral limit. This
phenomenon should not take place if the inter-instanton
interaction due to massless
quarks is present. Indeed, simulations of an ensemble of {\it interacting}
instantons do not show this enhancement at $\lambda = 0$.
A similar artificial peak at $\lambda = 0$ should be observed in quenched
lattice calculations, but it should be absent in simulations with dynamical
fermions. However, existing lattices are far too small to contain hundreds
of instantons and to reveal this phenomenon clearly.

In practice, the effect of this
enhancement is small due to the finite quark masses which smear out
the small eigenvalues. Considering the peak at $\lambda = 0$ as an artifact
of the quenched approximation we will compensate for this by taking
the quark masses somewhat larger than their phenomenological values, see
details below\footnote{In the case of dynamical quarks we have another
reason for not taking  the quark masses too small: at finite
volume the smallest eigenvalues are $\sim 1/V$ leading to a dip in the
spectrum at $\lambda = 0$.}.

\renewcommand{\theequation}{3.\arabic{equation}}
\setcounter{equation}{0}
\vskip 1.5cm
\centerline{\bf 3. Quark propagators: generalities}
\vskip 0.5cm

  In this section we discuss the quark propagator in a
multi-instanton gauge field.
Before going  into details, let us clarify the
following important point. Strictly speaking, the propagator
is not a physical quantity because it is {\it gauge-dependent}.
As discussed in detail in {\it e.g.} \cite{SHURYAK-1989III},
this problem may be
circumvented {\it \`a la} Schwinger by complementing its definition
with the path ordered exponential
$P\exp[(ig/2)\int_x^y A^a \sigma^a_\mu dx_\mu)]$.
Physically, this corresponds to the addition of a static anti-quark:
thus one naturally
proceeds to the discussion of heavy-light mesons (see below).

Although propagators are {\it not  physical quantities},
the specific feature of the 'instanton
vacuum' (which contains neither perturbative nor confining vacuum fields!)
is  that
this extra Schwinger factor is actually very close to one: static quarks
essentially {\it ignore} instantons. This statement was
demonstrated explicitly in \cite{SHURYAK-1989III}, and a related  analytical
argument was given in \cite{DIAKONOV-PETROV-POBYLITSA-1989}.
In this sense, one may say that our data on propagators
are 'practically gauge invariant', and therefore are not purely academic
quantities\footnote{Readers not satisfied by this argument should note
that the main objects of these
papers, the correlation functions, are manifestly gauge invariant by
construction.}.

The total spinor propagator
\be
S(x,y) = <x| \frac{-1}{i\gamma_\mu D_\mu + im}|y>,
\ee
will be approximated by the sum
\be
\sum_{I,J}<x|I><I| \frac{-1}{i\gamma_\mu D_\mu + im}|J><J|y> +
\sum_{P,Q}<x|P><P| \frac{-1}{i\gamma_\mu D_\mu + im}|Q><Q|y>,\nonumber\\
\ee
where $I$ and $J$ are zero modes, or the 'bound states' in terms of
the analogy put forward in chapter 2, and
$P$ and $Q$ are nonzero modes which can be thought of as
'scattering states'.
The latter can only be numerated by a continuum of eigenvalues,
{\it e.g.} the  momenta at large distances which makes
the complete diagonalization of the Dirac
operator a very complicated task. The simplifying assumption we have
made is that the cross term between both type of modes is small and
can be neglected. The first term follows from the Dirac operator in
the space of zero modes (see eq. (2.5)) and is given by
\be
S^{ZM}(x,y) = \phi_I(x) \left ( \frac 1{T -im} \right)_{IJ} \phi^*_J(y).
\ee
The second term is written as the sum of two contributions
\be
<x|\frac 1{-(\gamma\cdot D)^2 + m^2} i\gamma \cdot D|y>' +
im <x|\frac 1{-(\gamma\cdot D)^2 + m^2}|y>',
\ee
where the $\,'\,$ indicates that the zero mode contribution should be
excluded in the calculation of the expectation value.
In the second term we will neglect the interaction of the spin-induced
gluomagnetic moment with the field. This amounts to the replacement of
$(\gamma\cdot D)^2 \rightarrow D^2$, and is exact for
some correlators \cite{ANDREI-GROSS-1978}. What remains
is the propagator of a scalar quark which will be discussed in section 4.
The first term is the most complicated one and we postpone its discussion until
section 5.

The propagator in the approximation (3.1) still contains all possible
gamma matrix structures. In other words, if we write
\be  S &=&\sum a_i \Gamma_i \\
\Gamma_i&=&1, \quad\gamma_5,\quad \gamma_\mu,\quad i\gamma_5\gamma_\mu,\quad
 i\gamma_\mu\gamma_\nu\,\,\,
 (\mu \not= \nu),
\ee
all coefficients $a_i$, which are $SU(3)$-matrices, are nonzero.
However, only two coefficients
have a nonzero {\it average} value: $a_1$ and $a_{\gamma_0}$
(we assume that the propagation takes place in
time direction). On the other hand, the average of the combination
${\rm Tr}\, a_i a_i^\dagger \ne 0$ for all structures. In fact,
linear combinations of these quantities are nothing else
but mesonic correlation
functions which
will be discussed in \cite{SHURYAK-VERBAARSCHOT-1992D}.

The ensemble average of the propagator will be evaluated numerically via
a Monte-Carlo simulation of 256 instantons in a box of $3.36^3\times 6.72
fm^4$.
The distribution of the position of the instantons
will be taken uniform, whereas the size is kept fixed at $0.35 fm$. The
orientations are sampled from the invariant group measure. The average
propagators $<S(x+\tau,x)>$ are calculated by
averaging over an ensemble of 50 configurations
and over 100 randomly chosen initial points $x$ for each configuration and
for each value of the separation $\tau$ from the initial point.
To $u$ and $d$ quark masses are taken equal to 10 $MeV$, which is still
larger than the physical value. As discussed in section 2, the effect
of the anomalous large number of small eigenvalues is suppressed this way
(see Fig. 1). The strange mass is taken equal to the 'common sense' value
of 140 $MeV$,

\renewcommand{\theequation}{4.\arabic{equation}}
\setcounter{equation}{0}
\vskip 1.5cm
\centerline{\bf 4. Scalar quarks: propagators and  mesons}
\vskip 0.5cm

In this section we study the propagator of a massless scalar quark
(or {\it squark}, for brevity).
Its analytical expression for one instanton
is known\footnote{Here and below the instanton field is assumed to be
in the singular gauge, $A^a_\mu=2\bar\eta^a_{\mu\nu}x_\nu
\rho^2/[x^2(x^2+\rho^2]  $, where $\eta^a_{\mu\nu}$ is the 't Hooft
symbol.} \cite{BROWN-ETAL-1978}:
\be
D(x,y)= {1 \over 4\pi^2(x-y)^2} \frac 1{\sqrt{1+\rho^2/(x-z)^2}}
\left [1+\frac{\rho^2 \sigma^- \cdot (x-z) \,\sigma^+ \cdot (y-z)}
{(x-z)^2 (y-z)^2}\right ]
\frac 1{\sqrt{1+\rho^2/(y-z)^2}},\nonumber\\
\ee
where $z$ denotes the position of the center of the instanton.
The propagator in the field
of an anti-instanton is obtained by interchanging
$\sigma^+$ and $\sigma^-$.
 If the instanton is
rotated in color space by the matrix $R$, the propagator
should be rotated by the matrix $R^{ab}\sigma^b$.

  The first factor $D_0=1/ 4\pi^2(x-y)^2$
is the free scalar propagator.
When all distances involved exceed the instanton size $\rho$, the remaining
factors approach 1, leaving only the free propagator.
Slightly less trivial
is the short-distance limit $(x-y)^2 \rightarrow 0$.
Expanding in powers of $(x-y)^2$ we obtain
\be
D_I(x,y)={1 \over 4\pi^2(x-y)^2} +
{i\rho^2\eta_{\mu\nu}^a (x-z)_\mu (y-z)_\nu \sigma_a
\over 4\pi^2(x-y)^2 (x-z)^2 (y-z)^2}-
 {\rho^2 \over 8\pi^2 (x-z)^2 (y-z)^2}+ \cdots,
\ee
and as the numerator of the second last term is proportional
to $(x-y)$,
the free propagator $D_0$ dominates at small distances as well.

  The  very complicated problem of
 propagation in a multi-instanton environment
is significantly simplified at {\it not too large distances}, where only
one or a few close instantons contribute. Analyzing the
expression given above, one realizes that
significant corrections only appear if either
$x$ or $y$ or both are well {\it within } an instanton, {\it i.e.}
at a distance no more than
$\approx \rho$ from its center.
In \cite{SHURYAK-1982} only the effect of
the 'closest' instanton was taken into account.

One might think that one could improve on this by treating the effect
of the other instantons as small perturbations  and simply adding their
contributions,
\be
D_{\rm sc}=D_0+\sum_{I}(D_I-D_0).
\ee
As discussed in \cite{ANDREI-GROSS-1978}, this correction is the first term of
a 'rescattering' series.
   However, there is a serious problem: the integrated
contribution of {\it distant} instantons diverges because
of the term $\int d^4z /(x-z)^2 (y-z)^2$.

In order to say more about of the origin of this term let us look at the
expansion (4.2) in greater detail. When we allow for the possibility
of a nonzero current squark mass the expansion of
$1/(-D_\mu^{2}+m^2)$ in powers of the gauge field results in
\be
\frac 1{-D_\mu^2+m^2}(x,y) = D_m(x,y) +\int d^4 u D_m(x,u)
i\sigma^a A^a_\mu(u-z) \del_\mu D_m(u,y) + \cdots
\ee
This expansion which also was considered in \cite{LEVINE-YAFFE-1979}
provides us with the corrections of $O(\rho^2)$.
The interpretation of the correction
in this formula is that the quark propagates to some point $u$ in space time,
then exchanges 2 or more gluons with the center of the instanton (obtained
from the expansion of the instanton profile in inverse powers of $(u-z)^2$)
and finally proceeds to
its end point.
Therefore the factors $1/(x-z)^2$ and $1/(x-y)^2$, obtained after integration
over $u$  do not have an unambiguous
interpretation as the propagator of a gluon or of a squark. Both
gluons and squarks acquire a mass due to
non-perturbative effects, and therefore do not propagate far.
Writing the propagator
as
\be
D_I(x,y) = D_0(x,y) + \frac 1{4\pi^2(x-z)^2} \delta D \frac 1{4\pi^2(y-z)^2},
\ee
suggests a cure to our 'infrared problem':
one should substitute the massless propagators by the massive
propagators
 $D_M(x,z)$ and $D_M(y,z)$, which
makes the integral convergent\footnote{This  solution to our
infrared problem is in fact similar to the well known diagram
resummation in plasma physics \cite{GELLMANN-BRUECKNER-1957}, which takes care
of Debye screening.}. Since the divergence is logarithmic, the exact form of
the cutoff is not very important. The value of the mass
is certainly larger than the current quark masses. The natural
scale for this cut-off mass $M \approx \Lambda_{QCD}$.
In the case of the strange quark one might argue to add its mass
to the cutoff mass, but since we are already considering a relative
small correction, we will refrain from this kind of fine tuning.

Results  for ${\rm Tr}\,D(x,0)$ as a function of the
distance are shown in Fig. 2a. For convenience, ${\rm Tr}\,D(x,0)$
is normalized to the free massless
scalar propagator $D_0=1/4\pi^2x^2$, so the
deviation from 1 is caused by the deflection of the squark
in the instanton gauge field.
The solid
line corresponds to a {\it massive} scalar propagator, with
a fitted mass value of  $M_{\rm squark}= 140\, MeV $.
Note, that this value is roughly  half the
'constituent quark mass', expected for {\it spinor} quarks from
phenomenological considerations.

  In Fig. 2b we show the simplest {\it mesonic} correlation
function, obtained by
averaging the {\it  square} of the squark propagator
\be
K(x)= <{\rm Tr}(D(x,0)D(0,x))>.
\ee
As usual, the data shown are normalized with respect
to the perturbative correlator, and should be
close to 1 at small $x$.
The solid line corresponds to the {\it square} of the massive propagator,
fitted
to  data  in Fig. 2a. If two
quarks would propagate independently,
the data would be in agreement with it. However, all
 measured points are above this curve, which means that
scattering on instantons  not only generates
 an effective squark mass, but
also an  {\it attractive} interaction. The fitted value of the meson mass
is $\approx 150 MeV$, which is  less than $2 M_{\rm squark}\approx 280 MeV$.
Of course, it is natural that due to such attractive interaction
scalar quarks form mesons.

Note, that
the origin of this attraction at small distances can be traced to the second
term in the r.h.s. of (4.2):
its average ({\it e.g.} over orientations)  is zero,
but not the average of its square. It is the simplest example of 'hidden'
components of the quark propagators to be discussed
in the next paper of this
series \cite{SHURYAK-VERBAARSCHOT-1992D}.

 Concluding this section, we have shown that the
instanton vacuum can produce an effective squark
mass and an interaction between them.
This implies that similar effects can be expected for spinor quarks as well.

\renewcommand{\theequation}{5.\arabic{equation}}
\setcounter{equation}{0}
\vskip 1.5cm
\centerline{\bf 5. Effect of non-zero modes for spinor quarks}
\vskip 0.5cm

The main special feature of the spinor propagator is that it also receives
a contribution from the zero modes. However, before discussing the complete
propagator we first study the part of non-zero mode contribution
that  is given by the first term in eq. (3.4).
It is known analytically
for a {\it massless} quark in the field of a single instanton
\cite{BROWN-ETAL-1978}:
\be
S_I(x,y) &=&
\frac 1{\sqrt{1+\rho^2/x^2}} \frac 1{\sqrt{1+\rho^2/y^2}}\left (S_0(x,y)(
 1+\frac{\rho^2 \sigma^-\cdot x \,\sigma^+\cdot y}{x^2 y^2})\right .
\nonumber\\
&-& \left .D_0(x,y)\frac{\rho^2}{ x^2 y^2}
(\frac{\sigma^-\cdot x \,\sigma^+\cdot
\gamma \,\sigma^-\cdot \Delta \,\sigma^+\cdot y }{\rho^2+x^2}\gamma_5^+
+ \frac{\sigma^-\cdot x \,\sigma^+\cdot
\Delta \,\sigma^-\cdot \gamma \,\sigma^+\cdot y }{\rho^2+y^2}\gamma_5^-)
\right ) ,\nonumber\\
\ee
where we have introduced the projectors $\gamma_5^\pm = (1\pm\gamma_5)/2$.
The massless free quark propagator is denoted by $S_0(x,y)$, and the
free massless scalar propagator, $D_0(x,y)$ is given by the first term
in the r.h.s. of eq. (4.6).
The expression for anti-instantons is obtained by interchanging
$\sigma^+$ and $\sigma^-$. The $SU(3)-$extension of the propagator
is obtained by embedding this propagator in the upper $2\times 2$
block of a $3 \times 3$ matrix, and substituting the free propagator
in the remaining diagonal matrix element. For arbitrary orientation, $U$,
of the instanton $S_I(x,y)$ has to be rotated covariantly.

We generalize $S_I(x,y)$ to {\it many} instantons,
in the same approximate way as in the previous
section for squarks,
namely by {\it summing all deviations} from the free propagator:
\be
S_\gamma(x,y)=S_0(x,y)+\sum_I(S_{I}(x,y)-S_0(x,y)).
\ee
Although in this case the  contribution of distant instantons is not
divergent, we still think it should be excluded on the basis that
gluons cannot propagate that far.
If the NZM-induced effects are
treated as a correction, it is logical to also include here
an effective mass to cut off the contribution of distant instantons.

  The last necessary step is to include the effects
of  a non-zero {\it bare quark mass}: this is especially
needed for the discussion of correlators involving strange quarks.
  Generally speaking, there are at least three different scales
to which the strange quark mass, $m_s$, should be compared:\\
(i) The eigenvalues of the Dirac operator  in the ZMZ which are {\it
comparable} to $m_s$. Therefore, one should be careful
at this point and
not treat $m_s$ as a small parameter\footnote{Note that at this point
we deviate from what is done in chiral perturbation theory.}: the mass in
the strange
component of the denominator of the ZMZ term in the propagator
will be taken equal to $m_s$.\\
(ii) The field strength of the gauge field inside
instantons. Its value  is {\it very large}:
$\sqrt{G_{\mu\nu}^2}\sim 8\sqrt{3}/g\rho^2 \sim 2\, GeV^2 \gg m_s^2 \sim
.02\, GeV^2$, and therefore the  motion of strange quarks
inside instantons is essentially the same as that of massless ones.\\
(iii) The total (Euclidean) propagation time. In this work this time
is comparable to the strange quark mass ($\tau \sim 1-2 fm \sim m_s^{-1}$),
which makes the strange quark propagators
substantially different from non-strange ones. However, in a
reasonably dilute instanton
vacuum
most of this
motion takes place in relatively {\it empty} space, and one may therefore
describe these corrections using the free massive propagator.

  In such a complex situation, the problem is to work out a useful
{\it interpolating formula}, which combines several
limiting cases for which an analytical solution is known and presumably
provides a reasonable description in the general case. After trying several
expressions, we arrived at the following expression in which
factorization of the mass-related
damping of the propagator from corrections of scattering on instantons
is assumed:
\be
S_\gamma(m,x,y)=S_m(x,y) + D_M(x,0) \delta S(x,y) D_M(0,y).
\ee
The correction $\delta S$ is given by
\be
\delta S(x,y) = \frac {S_\gamma(x,y) - S_0(x,y)}{D_0(x,0) D_0(0,y)},
\ee
and $S_m(x,y)$ is the {\it massive} free fermion propagator.
At small distances (inside instantons) the the propagation of quarks
approaches free propagation,
while at large $|x-y|$ the quarks move as free massive fermions
in between successive scatterings  on instantons while
the correction $\delta S$ is exponentially suppressed.
We will choose $M = \Lambda_{QCD}$
which we believe to be a natural choice for this cut-off mass.

Finally we proceed to the expression for the total propagator
$S(x,y)$ by adding the contributions discussed is this section
and in section 3 and 4.
As a result we find
\be
S(x,y) = S_\gamma(m,x,y)+
\phi_I(x) \left ( \frac 1{T -im} \right)_{IJ} \phi^*_J(y)
+im D_{\rm sc}(x,y),
\ee
where $D_{\rm sc}$ is defined in eqs. (4.3) and (4.5) modified as discussed
below eq. (4.5).

In Fig. 3 we show the chirality flip ${\rm Tr}\, S$  (a) and the chirality
non-flip ${\rm Tr}\, \gamma_0 S$ (b) components of this propagator.
The squares represent results for the full propagator, whereas the crosses
are for the propagator $S(x,y) = S_m(x,y) + S^{ZM}(x,y)$, which includes
only the modification of the propagator due to zero modes.
The first trace ${\rm Tr}\, S$ is normalized with
respect to the short distance limit of the massive free
quark propagator ${\rm Tr}\, S_0(x,m)/m$ such that its value at $x = 0$
is equal to the mass $m$. The second trace is normalized with respect to
the free propagator.
One observes that their effect is significant, especially
at medium distances $x \sim 1/2 fm$.

The physical question we are going to discuss is whether or not
these results can be interpreted as the appearance of some
{\it universal} effective mass $m_{eff}$\footnote{The existence
of a momentum dependent  effective mass has also been obtained
analytically in the large $N_c-$limit \cite{DIAKONOV-PETROV-1986}.}.
For this reason
we have plotted the behaviour of the {\it massive} fermion propagator for
$m=200,\,300,\,400\,\, MeV$. At small $x$, the calculated points start
from zero, consistent
with bare quark propagation with a small input mass. However,
for $x>0.5 fm$ they even somewhat 'overshoot' the curves, but qualitatively
imitate their $x-$dependence. Since none of the curves is
particularly close to the data trend
it is not possible to select a 'preferable' mass from the chirality flip
amplitude.

The chirality non-flip component of the propagator, shown
in Fig. 3b, tends to 1 at small $x$
(which corresponds to massless propagation), but
at larger $x$ we do find qualitative
agreement with the massive free propagators.
The square points follow the
pattern given by the solid line.
This means  that
the net effect of the 'instanton vacuum' on quark propagation can indeed
be represented approximately by the
appearance of an effective mass of the order of $300\, MeV$.
This is about twice larger than for the scalar quark. Indeed, when
only the contribution of the zero modes is included, as shown by the
crosses in Fig. 3b,  the 'effective mass' is about twice smaller.
This shows that roughly half the constituent mass is due to the zero modes
and the other half due to the non-zero modes.

 Note that the curves in Fig. 3
imply that the amplitude of 'bare-to-dressed' quark transition amplitude
(the $Z$-factors introduced in \cite{SHURYAK-1989I}) is exactly 1, so
this feature of '100 percent dressing'
is approximately reproduced by our propagators as well.

In conclusion, there is some qualitative agreement with the
'constituent quark' model. However, one should not take  it
 literally. First of all, we remind that
agreement is
not really good for the chirality-flip amplitude shown in Fig. 3a.
And, last but not least, if two (or more) quarks propagate together,
they show a very strong interaction. Therefore,
as we will show in future publications, {\it none} of the mesonic or baryonic
correlators measured are actually reproduced by
 the naive 'constituent quark' model.

\renewcommand{\theequation}{6.\arabic{equation}}
\setcounter{equation}{0}
\vskip 1.5cm
\centerline{\bf 6. Heavy-light mesons}
\vskip 0.5cm

   Heavy-light mesonic correlators were first discussed by one
of us in \cite{SHURYAK-1982D}, where the concept
of 'heavy quark symmetry' between {\it any} channels with sufficiently heavy
quark masses was introduced.
Indeed, in the heavy quark limit $M_Q \gg
\Lambda_{QCD}$ the heavy quark behaves as a static center (the mesons'
'nucleus'), and the corresponding correlation functions essentially
reduce to the propagator of the light quark. As the direction of the spin
of heavy quark is irrelevant in this static limit, a specific degeneracy of
correlation functions follows. For example, the spatial component of the
vector correlator coincides with
pseudoscalar correlator, and the spatial component of the axial
correlator coincides with scalar correlator.

More specifically, we discuss correlators of currents
containing one heavy $Q =(c, b, t)$ and one light quark $q$ defined by
the expectation value
\be
K_\Gamma(x) = < \bar Q \Gamma q \bar q \Gamma Q>,
\ee
where $\Gamma$ is one of the possible gamma matrix structures. In this
section we consider ${ \bf 1}$ (S), $\gamma_5$ (P), $\vec \gamma$ (V) and
$\vec \gamma \gamma_5$ (A).
The heavy quark propagator that enters (6.1)
is given by the nonrelativistic quark propagator times a path ordered
exponential.
In \cite{SHURYAK-1989III} it was shown that the correction
due to the path ordered exponential is small.
This allows us to translate the propagator immediately into heavy-light
correlation functions. Using that only the large components of the
heavy quark spinor are relevant one obtains the correlator
\be
K_\Gamma(x)= i{\rm Tr}(S(-x) \Gamma {1+\gamma_0 x/|x| \over 2} \Gamma)
\ee
times a non-relativistic quark propagator. The separation $x$ is
chosen along the positive time axis.
Only the parity of the current is essential,
and for the parity $P = \pm 1$ channels with
$\Gamma=1,\,\gamma_5$ one obtains the following
correlation functions
\be K_{\pm}(x)= i{\rm Tr}[(\frac{1\pm\gamma_0}2 )S(-x)].
\ee
   As was noticed in \cite{SHURYAK-1982D}, the splitting of these
correlators at small distances, normalized to the free quark correlator
$K_0$,  is simply proportional to the quark condensate
\be K_\pm(x)/K_0(x) = 1 \mp {\pi^2 x^3 \over 6}|<\bar \psi \psi>| + \cdots.
\ee
  In order to describe the correlator in the whole region, we use a
standard parametrization for the spectral function,
a resonance plus the
perturbative continuum above a certain 'threshold' energy $E_0$, in this
case given by \cite{SHURYAK-1982D}:
\be {\rm Im} K_\pm(E)= 6\pi n \delta(E-E_{\rm res})
+ \theta(E-E_0){3 E^2 \over 2 \pi},
\ee
where $n=f_Q^2 M_q/12$ is the  3-dimensional density of the light quark at the
center. In terms of the space-time correlator, it translates into the following
expression:
\be
K_\pm(x)/K_0(x)= 2\pi^2 n x^3 e^{-E_{\rm res}x} +
(1+E_0 x + E^2_0 x^2/2)e^{-E_0 x}.
\ee
  Our results for the propagator, in the form $K_\pm(x)/K_0(x)$,
are shown in Fig. 4, where
the symbols $V$ and $A$ refer to the {\it spatial}
components of the vector and axial currents, respectively.
The dashed curves correspond to the three-parameter
fit (6.6), with the  values of the parameters given in Table 1.
For comparison we also give
results of previous works using various methods\footnote{
Comparison of compiled lattice results
 with results from QCD sum rules, as well as the discussion of
 $1/m_Q$ corrections can be found e.g. in \cite{ELETSKY-SHURYAK-1992}.}.

  The resonance position can also be compared to the experimental values for
$B$ meson masses, provided that the $b$ quark mass value is
obtained from different sources ($e.g.$
sum rules for upsilons). Assuming that $m_b \approx 4800$ MeV,
as follows from such sources, one obtains for
$E_{\rm res}=m_B-m_b \approx 475 MeV$, which is not too far from
our fitted value of 615 MeV.

The splitting of opposite parity states is only known for
charmed mesons (such states have not yet been discovered for
$B$ mesons): it is about 450 $MeV$ (see discussion is \cite{SHURYAK-1992}),
to be compared to our
estimated difference of 575 $MeV$. This value is very close to
twice the constituent quark mass, a
result which also was derived in \cite{NOWAK-ZAHED-1992} using
completely different methods.

It is  instructive to consider the small distance expansion of the
correlation function (6.6),
\be
K(x)/K_0(x)= 1 +
(2\pi^2 n - \frac 16 E_0^3) x^3 + (\frac 18 E_0^4 - 2\pi^2 n E_{\rm
res}) x^4 + \cdots.
\ee
Equating this to (6.4)
we find the so called 'duality relation',
\be
\pm \frac{\pi^2}{6} <\bar \psi\psi> = (2\pi^2 n^{\pm} - \frac 16 {E_0^\pm}^3),
\ee
which should be approximately satisfied. From the r.h.s. we find the
value for the quark condensate
 of  $-(215\,MeV)^3$ in case of the $0^-$ state and $-(578\,MeV)^3$
in case of the $0^+$ state. The discrepancy between these two numbers
is a signature of an early deviation of our fit and, possibly, the true
correlator from the predictions of the operator product
expansion. This also follows from the right hand side of the
dispersion relation which  can
be expanded in powers of $x$ for $x \ll 1/E_0\sim 0.2 fm$  only.

\vskip 1.5cm
\centerline{\bf 7. Conclusions and discussion}
\vskip 0.5cm

   In this work we report an extensive numerical
  study of the simplest ensemble of instantons: the random
instanton vacuum. This vacuum is characterized by two parameters:
(i) the size $\rho_0$ and
(ii) the density $N/V$ of instantons (plus anti-instantons) in the
QCD vacuum, which were chosen to reproduce the size of the quark
condensate and the gluon condensate.

The topological charge density in this model fluctuates strongly on all length
scales, which shows up in the form of an excess of small eigenvalues
of the Dirac operator
above the expected gaussian distribution.

For simplicity, we first studied the propagation of scalar quarks (or squarks),
in which case there is only a nonzero mode contribution to the propagator.
It appears that the scalar mesonic correlation function
can be described remarkably well by a 'constituent quark' model
with a mass of about 140 $MeV$.

The spinor quark propagator has two different components: the
chirality non-flip and the chirality flip parts. The first one can again be
described by an effective quark mass of $m_{\rm eff} \approx 300\, MeV$.
The second component is more complicated and corresponds approximately
to a free massive propagator at distances larger than about
1/2 $fm$. At distances of $x \approx 0.5-1.0 fm$
both the zero and non-zero fermion modes
contribute a comparable amount to the constituent quark mass.

  Observable consequences of these statements are most clearly seen
in the spectra of heavy-light mesons. We find a mass difference
of 615 $MeV$ between the $B-$meson and the bottom quark
which agrees well with the phenomenological value.
The splitting between the mass of opposite parity states was found to be
575 $MeV$, which is roughly equal to twice the constituent quark mass.

\vglue 0.6cm
{\bf \noindent Acknowledgements \hfil}
\vglue 0.4cm
 The reported work was partially supported by the US DOE grant
DE-FG-88ER40388. We acknowledge the NERSC at Lawrence Livermore where
most of the computations presented in this paper were performed.

\newpage
\vskip 1cm
\begin{tabular}{||l|l|l|l|l|l||} \hline
 channel $J^P$ & $E_{\rm res} [MeV]$ & $n=f^2 M_Q/12, [fm^{-3}]$
 &  $E_0$ [MeV] & ref. & comment    \\ \hline
$0^-,1^-$  & 615  & 1.2 &  1010 & this work & random instantons\\
   & 400$\pm$100  & 1$\pm$.5 & 900$\pm$100 & \cite{SHURYAK-1982D} &
QCD sum rules \\
   & 480$\pm$80  & 1$\pm$.3 & 700 & \cite{SHURYAK-1989III} &
IIA \\
   & 500$\pm$30  & 2.$\pm$.5 & 1100$\pm$100 & \cite{ELETSKY-SHURYAK-1992} &
sum rules \\
   &  - - -  & 4.5$\pm$1 & - - - & \cite{MARTINELLI-1991,MAIANI-1991} &
lattice, 'static'  \\
   &  - - -  & 2.5$\pm$1. & - - - & \cite{MARTINELLI-1991,MAIANI-1991} &
lattice,  'dynamic' \\
\hline
$0^+,1^+$  & 1190  & 4.6 &  1745 & this work & random instantons\\
   & 1200$\pm$200  & 6.3 & 1800$\pm$200 & \cite{SHURYAK-1982D} &
QCD sum rules \\
   & 1000$\pm$ 200  & .7$\pm$.5 & 1200$\pm$200 & \cite{SHURYAK-1989III} &
IIA \\
 \hline
\end{tabular}
\vskip 1cm
\centerline{\bf Table 1.}
\vskip 1cm
  Fitted parameters of heavy-light mesons, compared to those derived in
previous works. $E_{\rm res},\,n,\,E_0$ are the resonance mass
(counted from the heavy
quark mass), the density of the light quark at the center and the 'asymptotic
freedom threshold' (see text).
For comparison, we give some lattice results based on
a  static heavy quark and  a $1/m_Q$ expansion, as well as  results
based on the
extrapolation of simulations with dynamical quarks.

\newpage
\centerline{\bf Figure Captions.}
\vskip 0.5 cm
\noindent
Fig 1. The eigenvalue density $n(\lambda)$
of the Dirac operator in the space of zero modes for an ensemble of
512 and 128 instantons. The normalization $\int_{-\infty}^\infty  d\lambda
\rho(\lambda)= 1$
and the bin size used to obtain the histograms was 0.025. The variance of the
gaussian curve is equal to the variance of $n(\lambda)$ for $N=512$.
\vskip 0.5 cm
\noindent
Fig. 2. The average propagator of a scalar quark (a) and its averaged square
(b)
(corresponding to the scalar meson), normalized to the propagator of a massless
quark. The solid lines in (a) and (b) correspond to
a propagator with a  fitted mass value of 140 $MeV$.
\vskip 0.5 cm
\noindent
Fig. 3. The chirality-flip (a) and non-flip
(b) components of the quark propagator
versus the distance $x$ (in $fm$). The normalization,indicated in the figure,
is discussed in the text. Crosses in (b) correspond to the effects of
zero-modes
only, while the squares give the complete result. Three lines, the
short-dashed, solid and long-dashed correspond to a massive free propagator
with a mass of 200,300 and 400 $MeV$, respectively.
\vskip 0.5 cm
\noindent
Fig. 4. The  correlation functions for negative
($P$ = pseudoscalar, $V$ = vector (spatial)) and positive ($S$
 = scalar, $A$ = axial (spatial))
parity  heavy-light mesons. The data points are our results, while the dashed
curves are fits, discussed in the text and in Table 1.
\pagestyle{empty}
\vfill
\eject

\vskip 0.5cm
Fig. 1
\vskip 0.5cm
Fig. 1
\vskip 0.5cm
Fig. 2
\vskip 0.5cm
Fig. 2
\vskip 0.5cm
Fig. 3
\vskip 0.5cm
Fig. 3
\vskip 0.5cm
Fig. 4
\vskip 0.5cm
Fig. 4
\vskip 0.5cm
Fig. 5
\vskip 0.5cm
Fig. 6
\vskip 0.5cm
Fig. 7
\vskip 0.5cm
Fig. 8
\vskip 0.5cm
Fig. 9
\end{document}